\pdfoutput=1
\documentclass[twocolumn,superscriptaddress,aps,preprintnumbers,amsmath,amssymb,prd,nofootinbib]{revtex4}

\usepackage{amsmath}
\usepackage{amssymb}
\usepackage{amsfonts}
\usepackage{graphicx}
\usepackage{xcolor}
\usepackage{xfrac}
\usepackage{comment}
\usepackage{pifont}
\usepackage{physics}
\usepackage{fourier}
\usepackage{hyperref}
\usepackage{bm}
\usepackage{enumitem}

\definecolor{rossoferrari}{HTML}{D9073D}
\definecolor{mediumblue}{HTML}{0000CD}
\definecolor{forestgreen}{HTML}{228B22}
\definecolor{desy_blue}{HTML}{009EE2}
\definecolor{desy_orange}{HTML}{FD8800}
\definecolor{light_pink}{rgb}{1,0.4,0.4}
\definecolor{light_blue}{rgb}{0.284602,0.317763,0.963947}
\hypersetup{setpagesize=false,bookmarksnumbered=true,bookmarksopen=true,%
colorlinks=true,linkcolor=light_blue,urlcolor=rossoferrari,citecolor=rossoferrari,linktocpage=false}

\def\d{{\rm d}}

\usepackage{slashed}

\def\cN{{\cal N}}

\def\cT{{\cal T}}
\def\cU{{\cal U}}

\def\bR{{\mathbb R}}

\def\bZ{{\mathbb Z}}

\def\so{{\mathsf o}}
\def\sp{{\mathsf p}}

\def\su{{\mathsf u}}

\def\Spin{\mathrm{Spin}}

\def\u{\mathfrak{u}}
\def\su{\mathfrak{su}}

\def\so{\mathfrak{so}}
\def\sp{\mathfrak{sp}}

\def\beq#1\eeq{\begin{align}#1\end{align}}

\def\hete{(E_8 \times E_8) \rtimes \bZ_2}
\def\hets{\Spin(32)/\bZ_2}

\def\uu{u}
\def\vv{v}
\def\uu{u}
\def\ww{w}

\begin{document}


\preprint{TU-1227}


\title{Dualities among Neveu-Schwarz sector branes in string theory}


\author{Kazuya~Yonekura}
\affiliation{Department of Physics, Tohoku University, Sendai, Miyagi 980-8578, Japan}


\begin{abstract}
\noindent

We conjecture dualities among branes that consist of Neveu-Schwarz sector fields, such as NS5-branes. Two branes are dual after compactifications on spheres with gauge fluxes. The conjecture follows from simple considerations on string worldsheet theories. For some NS5-branes, we get level-rank dualities of Chern-Simons theories. Application of the dualities to a newly found brane in heterotic string theory reveals surprising properties of the worldvolume theory on it, for which we give evidence from anomaly inflow. 

\end{abstract}


\maketitle


D-branes and their dualities such as T-duality have been well-investigated.
String theory also contains branes that consist purely of Neveu-Schwarz (NS) sector fields. NS5-branes~\cite{Callan:1991at} are most famous, but there are also nonsupersymmetric branes in $\hete$ and $\hets$ heterotic string theories~\cite{Polchinski:2005bg,Bergshoeff:2006bs,Kaidi:2023tqo}. 
The purpose of this paper is to propose dualities between a $p$-brane compactified on $S^{8-q}$ and a $q$-brane compactified on $S^{8-p}$, with appropriate gauge fluxes on the spheres.

A $p$-brane may be described by supergravity solutions. (For a review see \cite{Aharony:1999ti}.)
In the extremal limit, the metric and dilaton of an NS sector brane are of the form~\cite{Horowitz:1991cd,FKWY}
\beq
\d s^2 = \d x^\mu \d x_\mu + \d \uu^2 + R(\uu)^2 \d \Omega^2_{8-p}, \quad \Phi = \Phi(\uu),
\eeq
where $x^\mu~(\mu=0,\cdots, p)$ are coordinates parallel to the $p$-brane, $ \d \Omega^2_{8-p}$ is the metric of the sphere $S^{8-p}$, and $R(\uu)$ and $\Phi(\uu)$ are functions of the coordinate $\uu \in  \bR$. They behave as 
\beq
(R(\uu), \Phi(\uu)) \to \left\{ \begin{array}{ll} 
(\uu, ~\Phi_0) & \uu \to +\infty \\
(R_0,\, - c_0 \uu) & \uu \to -\infty
\end{array} \right.
\eeq
where $\Phi_0$, $R_0$ and $c_0$ are constants. The region $\uu \to +\infty$ corresponds to the flat space away from the brane, while $\uu \to -\infty$ is the near horizon limit and is called the throat region. There is also some gauge flux on $S^{8-p}$. 

The throat region may be a holographic dual of the worldvolume theory on the brane~\cite{Aharony:1998ub}. This region has the following worldsheet description. The directions $x^\mu$ are described by a free conformal field theory (CFT), and we denote it as $\bR^{p+1}_{\rm flat}$. The sphere $S^{8-p}$ of radius $R_0$ with the given gauge flux is described by a CFT which we denote as $\cT_0$.
The direction $\uu$ with the dilaton $\Phi = -c_0 \uu$ is described by a linear dilaton CFT which we denote as $\bR^{\rm dilaton}_{c_0}$. In the case of heterotic string theories, the worldsheet theory has an internal CFT $\cU$. If the gauge flux on $S^{8-p}$ involves gauge fields coming from $\cU$, the $\cU$ is replaced  by another CFT $\cU'$ (see \cite{Kaidi:2023tqo} for details).
 The total worldsheet theory is thus given by
\beq
\bR^{p+1}_{\rm flat} \times \bR^{\rm dilaton}_{c_0} \times \cT_0 \times \cU'. \label{eq:throatCFT}
\eeq

We consider a compactification of \eqref{eq:throatCFT} as follows. We assume $(8-q)+1\leq p$. Among $p$ space coordinates of $\bR^{p+1}_{\rm flat}$, we use $(8-q)$ coordinates for the sphere $S^{8-q}$ with a given gauge flux. This is described by a CFT which we denote as $\cT_1$. To match the central charge, we also need to take another linear dilaton configuration which we denote as $\bR^{\rm dilaton}_{c_1}$. This direction is taken from one of the coordinates of $\bR^{p+1}_{\rm flat}$. Depending on the gauge flux on $S^{8-q}$, the $\cU'$ may be replaced by another $ \cU''$. The worldsheet theory is
\beq
\bR^{p+q-8}_{\rm flat} \times \bR^{\rm dilaton}_{c_0} \times \bR^{\rm dilaton}_{c_1}  \times \cT_0 \times \cT_1 \times  \cU''.\label{eq:lowtheory2}
\eeq
Here, we have used the special property of NS sector configurations that a product of CFTs is a CFT. This is the point that is not straightforwardly applicable to configurations with Ramond-Ramond (RR) fluxes.

The above CFT can be simplified as follows. The dilaton for $\bR^{\rm dilaton}_{c_0} \times \bR^{\rm dilaton}_{c_1} $ is given by $\Phi = -c_0 \uu - c_1 \vv$ for the coordinates $(\uu,\vv)$ of $\bR^2$. By introducing $c = (c_0^2+c_1^2)^{1/2}$ and $\ww = (c_0 \uu + c_1 \vv)/c$, it becomes $\Phi = - c \ww$. Thus, we just get $\bR^{\rm dilaton}_{c_0} \times \bR^{\rm dilaton}_{c_1} =  \bR_{\rm flat} \times \bR^{\rm dilaton}_c$ and hence \eqref{eq:lowtheory2} becomes
\beq
\bR^{p+q-7}_{\rm flat} \times \bR^{\rm dilaton}_{c}  \times \cT_0 \times \cT_1 \times  \cU''. \label{eq:lowtheory}
\eeq
We assume that \eqref{eq:lowtheory} is the holographic dual of the $(p+q-7)$-dimensional theory obtained by compactifying the $p$-brane on $S^{8-q}$. The decoupling of the $p$-brane worldvolume theory from bulk dilaton values~\cite{Seiberg:1997zk} is a hint for it.

Now, suppose that we have a $q$-brane whose holographic dual is given by the sphere $S^{8-q}$ with the given gauge flux. 
In completely the same way as above, the compactification of the $q$-brane on $S^{8-p}$ with the given gauge flux leads to exactly the same CFT \eqref{eq:lowtheory}. 
This is the reason for the conjectural dualities mentioned at the beginning.  

Precise energy scales for the validity of the dualities 
are not immediately clear. In the following, we focus on gapless degrees of freedom in $(p+q-7)$-dimensions. However, it might be possible that the dualities are valid at higher energies as in the case of T-duality of D-branes.

Let us first consider the dualities for NS5-branes in Type~IIB string theory, i.e. $p=q=5$. NS5-branes in Type~IIB string theory are S-dual to D5-branes and hence the worldvolume theory on $n$-coincident NS5-branes is given by the six dimensional $\cN=(1,1)$ Super-Yang-Mills theory with the gauge algebra $\u(n)$. The $\u(1)$ part and the global structure of the gauge group are made complicated due to RR-fields, so we neglect them and focus on the nonabelian $\su(n)$ algebra, leaving more precise studies for future work. Let $H_3=\d B_2$ be the 3-form field strength of the NS $B$-field, and let $A_{\su(n)}$ be the $\su(n)$ gauge field. The worldvolume contains a term
\beq
\frac{1}{4\pi} \int H_3 \wedge  \tr \left( A_{\su(n)} \d A_{\su(n)} + \frac23 A_{\su(n)}^3 \right).
\eeq
If we compactify this theory on $S^3$ with a flux $\int_{S^3} H_3 = m$, this term gives Chern-Simons level $m$ to the $\su(n)$ gauge field in three dimensions. According to our conjecture, this is dual to the compactification of $m$-coincident NS5-branes on $S^3$ with a flux $\int_{S^3} H_3=n$. This gives $\su(m)$ Chern-Simons theory at level $n$. These two theories are dual by the level-rank duality \cite{Naculich:1990pa} (neglecting details on the abelian part.) 

If we consider NS5-branes in $\hets$ heterotic string theory, we get the level-rank duality between $\sp(n)$ at level $m$ and $\sp(m)$ at level $n$~\cite{Naculich:1990pa}.
For Type IIA and $\hete$ heterotic string theories, the low energy theories after compactification to three dimensions may be gapped and the dualities may become empty in the strict low energy limit.                

Assuming the validity of the dualities, the most interesting application may be to the branes found in \cite{Polchinski:2005bg,Bergshoeff:2006bs,Kaidi:2023tqo} in heterotic string theories, because the worldvolume theories on these branes are mysterious. 
Here we consider the duality between the 6-brane~\cite{Kaidi:2023tqo} and the NS5-brane in $\hets$ heterotic string theory. The 6-brane is characterized by the following $\hets$ gauge flux on $S^2$. We take $\u(1) \times \su(16) \simeq \u(16) \subset \so(32)$, and normalize $\u(1)$ in such a way that the fundamental (vector) representation of $\so(32)$ decomposes into charge $\pm 1$ representations of $\u(1)$. Let $f$ be the field strength of this $\u(1)$. On $S^2$, we put a half-integral flux $\int_{S^2} f/2\pi = 1/2$ which is allowed by the fact that the global structure of the group is $\hets$. 
The compactification of the 6-brane on $S^3$ with $\int_{S^3} H_3 = n$ is dual to the compactification of $n$-coincident NS5-branes on $S^2$ with the above gauge flux of $\hets$.

The worldvolume theory on $n$-coincident NS5-branes is the same as that of D5-branes in Type~I string theory. It contains $\sp(n)$ gauge field. There are also chiral fermions $\Psi_{+}^{\rm S}, \Psi_{-}^{\rm AS}$ and $\Psi_{-}^{\rm BF} $ where the subscripts $\pm$ represent chirality. The $\Psi_{+}^{\rm S}$ and $\Psi_{-}^{\rm AS}$ are in the second symmetric (or equivalently adjoint) and antisymmetric representations of $\sp(n)$, respectively. The $\Psi_{-}^{\rm BF} $ is Majorana-Weyl in the bifundamental representation of $\sp(n) \times \so(32)$.

For the gauge flux on $S^2$ to be consistent with Dirac quantization conditions, we also introduce a flux of $\sp(n)$ as follows. We take $\u(1) \times \su(n) \simeq \u(n) \subset \sp(n)$, normalize $\u(1)$ in a similar way to that of $u(1) \times \su(16) \subset \so(32)$, and put a half-integral flux $\int f'/2\pi = 1/2$ for the $\u(1)$ gauge field $f'$.  Zero modes on $S^2$ give four-dimensional chiral fermions $\psi_{+}^{\rm 2  S}, \psi_{-}^{\rm 2 AS},\psi_{-}^{\rm BF} $ where the subscripts represent chirality in four dimensions. The $\psi_{+}^{\rm 2 S} $and $\psi_{-}^{\rm 2 AS}$ are two copies of fermions in the symmetric and antisymmetric representations of $\su(n)$, respectively. The $\psi_{-}^{\rm BF}$ is in the bifundamental representation of $ \su(n) \times \su(16)$. (The $\u(1)$ part and global structure are neglected as before.) 

Thus, assuming the validity of the duality, the compactification of the 6-brane on $S^3$ with $\int_{S^3} H_3=n$ gives a gauge theory containing (among other things) $\su(n)$ gauge field and chiral fermions $\psi_{+}^{\rm 2  S}, \psi_{-}^{\rm 2 AS},\psi_{-}^{\rm BF} $. It is surprising that chiral fermions appear after compactification on odd-dimensional sphere $S^3$. Moreover the low energy gauge group depends on the 3-form flux $n$. It would be very hard to realize these properties in usual Lagrangian field theories.

Anomaly inflow gives a partial check of the matter content. $\hets$ heterotic string theory contains a term $2\pi \int B_2 \wedge \frac{1}{4!} \tr (F_{\so(32)}/2\pi)^4+\cdots$, where $F_{\so(32)}$ is the $\hets$ gauge field strength and $B_2$ is the $B$-field. If we compactify this term on $S^3 \times S^2$ with the above gauge fluxes, we get a five-dimensional term of the form
\beq
2 \pi   \int \frac{n}{3! (2\pi)^3} \tr \left[ A_{\su(16) }  (\d A_{\su(16) } )^2   +\cdots \right],
\eeq
where $A_{\su(16)} $ is the $\su(16)$ gauge field. This is precisely the term required to cancel the $\su(16)$ anomaly of the fermions $\psi_{-}^{\rm BF} $ via anomaly inflow. This kind of argument supports the general idea that \eqref{eq:lowtheory} describes the theories obtained after compactifications of branes. It would also be interesting to study more subtle anomalies along the lines of \cite{Tachikawa:2024ucm}.


The work of KY is supported in part by JST FOREST Program (Grant Number JPMJFR2030, Japan) and
MEXT-JSPS Grant-in-Aid 
(No. 21H05188 and 17K14265).



\bibliographystyle{JHEP}
\bibliography{ref}


\end{document}